# Superfunctional high-entropy alloys and ceramics by severe plastic deformation

Parisa Edalati[1,2,*], Masayoshi Fuji[1,2,*] and Kaveh Edalati[3,4,*]

[1] Department of Life Science and Applied Chemistry, Nagoya Institute of Technology, Tajimi 507-0071, Japan
[2] Advanced Ceramics Research Center, Nagoya Institute of Technology, Tajimi 507-0071, Japan
[3] WPI International Institute for Carbon-Neutral Energy Research (WPI-I2CNER), Kyushu University, Fukuoka 819-0395, Japan
[4] Mitsui Chemicals, Inc.—Carbon Neutral Research Center (MCI-CNRC), Kyushu University, Fukuoka 819-0395, Japan

High-entropy alloys and ceramics containing at least five principal elements have recently received high attention for various mechanical and functional applications. The application of severe plastic deformation (SPD), particularly the high-pressure torsion (HPT) method, combined with the CALPHAD and first-principles calculations resulted in the development of numerous superfunctional high-entropy materials with superior properties compared to the normal functions of engineering materials. This article reviews the recent advances in the application of SPD to developing superfunctional high-entropy materials. These superfunctional properties include (i) ultrahigh hardness levels comparable to the hardness of ceramics in high-entropy alloys, (ii) high yield strength and good hydrogen embrittlement resistance in high-entropy alloys; (iii) high strength, low elastic modulus, and high biocompatibility in high-entropy alloys, (iv) fast and reversible hydrogen storage in high-entropy hydrides, (v) photovoltaic performance and photocurrent generation on high-entropy semiconductors, (vi) photocatalytic oxygen and hydrogen production from water splitting on high-entropy oxides and oxynitrides, and (vii) $CO_2$ photoreduction on high-entropy ceramics. These findings introduce SPD as not only a processing tool to improve the properties of existing high-entropy materials but also as a synthesis tool to produce novel high-entropy materials with superior properties compared with conventional engineering materials.
***Keywords:*** multi-principal element alloys (MPEAs); high-entropy alloys (HEAs); high-entropy ceramics (HECs); high-entropy oxides (HEOs); ultrafine-grained (UFG) microstructure; high-pressure torsion (HPT).

* Corresponding authors: P. Edalati (E-mail: parisaedalati@gmail.com; Tel: +81-57-224-8110)
M. Fuji (E-mail: fuji@nitech.ac.jp; Tel: +81-57-224-8110)
K. Edalati (E-mail: kaveh.edalati@kyudai.jp; Tel: +81-92-802-6744)

## 1. Introduction

There have been significant activities within the past three decades to produce bulk ultrafine-grained (UFG) materials using severe plastic deformation (SPD) methods such as high-pressure torsion (HPT), equal-channel angular pressing (ECAP), accumulative roll-bonding (ARB), twist extrusion (TE), and multidirectional forging (MDF) [1-4]. The formation of UFG structures with a large density of lattice defects leads to exceptional mechanical and functional properties of these severely deformed materials [5-8]. Since the properties of these materials are superior to conventional engineering materials, a recent publication used the term superfunctional materials for them [9]. Although SPD was well-known in western countries several decades ago [10-12], the method owes its popularity to a few interesting works done by Russian scientists in the 1980s [13,14].

A special issue edited by Edalati and Horita in 2019 reviewed the historical developments and recent advances in the SPD field by inviting the most active scientists in the field [15,16]. These review papers covered a wide range of progress from the mechanism and theoretical aspects of SPD [17-20] to the processing aspects including traditional SPD methods [20-23] continuous methods [24], upscaled methods [25,26], *in situ* methods [19,27], and surface treatment methods [28,29]. Although SPD was defined as a metal processing method at the beginning of this century, its application has been currently extended to non-metallic materials by several research groups [19,27,30-33]. These review papers clearly confirm that the SPD methods are widely used by researchers from many countries to control microstructural features/defects [34-38] and solid-state phase transformations [39-42] to achieve a wide range of mechanical [43-48] and functional properties [32,49-56].

In recent years, many publications reported the application of the SPD process to high-entropy or multi-principal element materials, although these publications were not reviewed in the above-mentioned special issue. High-entropy materials, including high-entropy alloys [57-59] and high-entropy ceramics [60-62], which were first introduced in 2004 [63,64], are compounds containing at least five principal elements (5-35 at%) with a configurational entropy higher than 1.5$R$ ($R$: gas constant). High-entropy materials show promising properties mainly because of the cocktail effect, lattice distortion, sluggish diffusion, and high configurational entropy [57-62]. The SPD methods were successfully used not only as a processing tool to enhance the properties of high-entropy alloys but also as a synthesis tool to develop new high-entropy alloys and ceramics [65].

A summary of publications on the application of SPD to medium-entropy and high-entropy materials is given in Table A1 of the Appendix [66-169]. While most studies focused on microstructural features and mechanical properties, there are limited studies on functional properties. This article reviews the recent advances in the application of SPD to achieve superfunctional high-entropy materials.

## 2. Ultrahigh hardness

The most investigated properties of high-entropy alloys are mechanical properties [5,6], and it is believed these materials can have a higher hardness than conventional engineering alloys [170,171]. The high hardness is usually attributed to lattice strain and significant solution hardening in these alloys [170-173]. The introduction of these alloys resulted in some attempts to design metallic alloys with hardness levels comparable to brittle ceramics. A few studies designed new high-entropy alloys and processed them by HPT to achieve some of the highest hardness levels ever reported in metallic alloys. Carbon-doped AlTiFeCoNi is an alloy designed by using the thermodynamic calculations using CALPHAD (calculation of phase diagram) to have dual phases with some small amounts of carbide precipitate, as shown in Fig. 1a [136]. The alloy shows an ultrahigh hardness of 950 Hv after SPD processing, as shown in Fig. 1b. Such a high hardness was not only due to the high solution hardening effect but also due to some other hardening mechanisms such as the formation of nanograins with sizes smaller than 100 nm, interphase hardening shown in Fig. 1c, dislocation accumulation shown in Fig. 1d, and precipitation hardening [136]. The formation of nanograins with sizes smaller than 100 nm, which cannot be easily achieved in severely deformed metals and conventional alloys, should be due to the significant interaction of solute atoms with dislocations and grain boundaries in high-entropy alloys.

In another study, one of the highest hardness levels reported for metallic alloys (1030 Hv) was achieved in a newly designed dual-phase (cubic + hexagonal) AlCrFeCoNiNb alloy, as shown in Fig. 1e [17]. The introduction of 10 nm nanograin sizes (Fig. 1f), dislocations (Fig. 1g), interfaces, and spinodal decomposition were responsible for such a high hardness [137]. In summary, the SPD process can be used to produce high-entropy alloys with ultrahigh hardness levels comparable to ceramics, while the plasticity of alloys is naturally higher than ceramics. Here, it should be noted that processing of such ultrahard alloys is currently possible only by HPT. Since a disc or ring is torsionally strained under very high pressures in HPT [174,175], the method is applicable to hard and less ductile materials including metals (Hf [176], Mo [177], and W [178]), glasses [179,180], silicon [181,182] and even diamond [183,184]. Another issue that needs to be considered is that the authors' attempt to apply SPD to existing single-phase [166] and dual-phase [163] high-entropy alloys led to hardness values that were not comparable to those of carbon-doped AlTiFeCoNi and AlCrFeCoNiNb, and thus, the design of alloys to have the simultaneous contribution of different hardening mechanisms is essential for achieving ultrahigh hardness after SPD [136,137].

These studies clearly show that the combination of various strengthening mechanisms in high-entropy alloys by SPD processing is a new strategy to realize metallic materials with high hardness levels comparable to the hardness of ceramics. Although metallic materials are expected to exhibit higher plasticity compared to ceramics, it is essential to conduct future research on the enhancement of the ductility of these ultrahard high-entropy alloys to have a good combination of ultrahigh hardness and high ductility.

## 3. Hydrogen embrittlement resistance

With the quick development of hydrogen as a clean fuel, demand for materials that show high strength in the presence of hydrogen has increased [185]. The main drawback in this regard is that high-strength alloys, such as steels with martensitic or body-centered cubic (BCC) structures, show poor plasticity in the presence of hydrogen due to different hydrogen embrittlement mechanisms such as hydrogen-enhanced localized plasticity [186] or hydrogen-enhanced decohesion [187]. Although all materials basically suffer from hydrogen embrittlement, the alloys with face-centered cubic (FCC) crystal structures like austenitic steels have more compatibility with hydrogen [188,189]. Recently, high-entropy alloys with the FCC crystal structure, particularly CrMnFeCoNi Cantor alloy, were introduced as new hydrogen-compatible materials [190,191]; however, these alloys have low yield strength.

A recent study employed the HPT method to enhance the yield strength of Cantor alloy and examine its hydrogen embrittlement resistance after applying various levels of strains (i.e. after various HPT turns), as shown in Fig. 2a [167]. Although coarse-grained material exhibits a large elongation of about 80% in the presence of hydrogen, its yield stress is 220 MPa which is not a high strength for many engineering applications. When low levels of strain are applied to the material such as 1/16 rotations of HPT, large fractions of twins are formed (Fig. 2b), leading to more than 1 GPa yield stress in the presence of hydrogen with an appreciable elongation of about 9%. With a further increase in the applied strain such as 1/8-1/4 rotations of HPT, dislocation-based defects, particularly low-mobility Lomer-Cottrell locks (Fig. 2c) and D-Frank partial dislocations (Fig. 2d), are generated which can further enhance the yield strength, while retaining acceptable plasticity. With a significant increase in strain such as 1-10 HPT rotations, nanograins are formed leading to up to 1.9 GPa strength but with very poor resistance to hydrogen embrittlement due to the possible occurrence of hydrogen-enhanced decohesion at grain boundaries.

Another study applied SPD to the surface through surface mechanical attrition treatment (SMAT) and reported a combination of high yield stress (0.5-0.7 GPa) and high plasticity (15-33%) in the presence of hydrogen (Fig. 2e) [165]. Such impressive results were attributed to the formation of gradient microstructure with a large fraction of twins (Fig. 2f). These studies suggested that twins and low-mobility Lomer-Cottrell locks and D-Frank partial dislocations are appropriate defects in the Cantor alloy to obtain large yield strength and good plasticity due to the suppression of hydrogen-enhanced localized plasticity [165,167]. However, the formation of nanograins leads to poor hydrogen embrittlement resistance of the Cantor alloy due to fast grain-boundary hydrogen diffusion and domination of hydrogen-enhanced decohesion [165,167]. Although poor ductility is a well-known feature of many metallic nanomaterials which needs to be improved using special strategies [192-196], this effect seems to be more severe in the presence of hydrogen in high-entropy alloys leading to failure even in the elastic region [165,167].

These studies suggest that high-entropy alloys have a high potential to achieve a combination of high strength and high hydrogen embrittlement resistance by controlling the lattice defects through SPD processing. However, the strength level and resistance to embrittlement can be still improved by controlling the chemical composition of these alloys to suppress the diffusion of hydrogen through the surface to bulk as well as through the lattice. Moreover, a combination of the SPD process and heat treatment can be used for further engineering of lattice defects and controlling their interaction with hydrogen.

## 4. Biocompatibility

Severely deformed materials, particularly titanium and its alloys, can exhibit a good combination of high strength and good biocompatibility [197-199], and this has led to the commercialization of these materials for biomedical applications [200]. However, for materials used as implants for orthopedics, besides high strength and biocompatibility, the elastic modulus should be low and ideally comparable to the elastic modulus of human bone (10-30 GPa) [201]. Recently, high-entropy alloys were introduced as a new family of biocompatible materials with reasonable elastic modulus [202-205], but the strength of these materials needs to be enhanced by appropriate hardening mechanisms [206-210].

In a study, a lattice-softened Ti-containing high-entropy alloy, TiAlFeNiCo, was designed with the aid of thermodynamic calculations and further processed by SPD through the HPT method to enhance its strength [123]. The material had a coarse-grained structure with the BCC + $L2_1$ crystal structure after arc melting, but it showed phase transformation to the BCC phase after HPT processing together with the formation of UFG structure. These structural and microstructural modifications by HPT processing resulted in the enhancement of hardness and the reduction of elastic modulus to about 120 GPa, as shown in the plots of nanoindentation load against displacement in Fig. 3a. A comparison between the strength (examined by Vickers method) and cell proliferation-viability-cytotoxicity activity (examined by MTT assay) of this high-entropy alloy with those achieved for pure titanium and the Ti-6Al-7Nb (wt%) alloy is shown in Fig. 3b. The TiAlFeCoNi alloy shows 170-580% larger microhardness and 260-1020% higher cellular metabolic activity than pure titanium and the Ti-6Al-7Nb alloy, while its elastic modulus is similar to these two biomaterials [123].

Biocompatible high-entropy alloys can be also synthesized from the powders of pure metals by using the concept of ultra-SPD [134]. In ultra-SPD, the shear strain is significantly enhanced (usually over 1,000) so that the thickness of sheared phases becomes geometrically comparable to atomic distances [211]. Ultra-SPD is quite effective to synthesize a wide range of alloys even from immiscible systems [212-214] not only because of the mechanical strain effect but also due to ultrafast dynamic diffusion [215-217]. Among various SPD methods, HPT is currently the best method to impart ultra-SPD because the method can induce very large strains (only by continuous rotation of HPT anvils) without any contamination or significant temperature rise [218,219]. Therefore, the HPT method can be used to simultaneously conduct mechanical alloying, consolidate the powders at low temperatures and refine the grain size to the UFG region [212-214]. A recent study directly synthesized several biomaterials from the powders of pure metals by the concept of ultra-SPD: a binary TiNb alloy, a ternary TiNbZr alloy, a medium-entropy alloy TiNbZrTa, and a high-entropy alloy TiNbZrTaHf [134]. All synthesized biomaterials had single BCC phases with grain sizes at the nanometer level. As shown in Fig. 3c, the hardness increases with increasing the number of principal elements, while the high-entropy alloy with five principal elements exhibits the highest hardness. As shown in Fig. 3d, these medium-entropy and high-entropy alloys have elastic modulus values close to 80 GPa, which is comparable to or even better than many commercial biomaterials, while their hardness levels are almost the highest for the biocompatible alloys (> 500 Hv). In conclusion, a combination of the concepts of SPD and high-entropy materials appears as a promising approach to developing biomaterials with high strength, low elastic modulus, and good biocompatibility [123,134]. Corrosion resistance, deformation ductility, fatigue properties, and tribological properties are other critical parameters that need to be investigated for the practical application of SPD-processed high-entropy materials for biomedical applications [220], although a study reported good corrosion resistance of these alloys after SPD processing [150].

A good combination of high strength and high biocompatibility of severely deformed high-entropy alloys compared to conventional biomaterials introduces them as a new family of biomaterials. However, the future direction should be the design of alloys with appropriate density and a lower elastic modulus. Moreover, although aluminum is still used in biomedical alloys such as Ti-6V-4Al and Ti-6Al-7Nb (wt%), this element should be excluded from the composition of conventional and high-entropy alloys in the future because of biomedical toxicity issues.

## 5. Hydrogen storage

Hydrogen is considered a clean fuel for the future, but since hydrogen is a light gas, its safe and high-density storage is a big academic and industrial challenge [221]. Metal hydrides provide a promising technology for compact and low-pressure storage of hydrogen [222,223]; however, many hydrides suffer either from kinetic issues (i.e., activation and hydrogenation/dehydrogenation rate) or thermodynamic issues (i.e., the high stability of hydride at room temperature) [224,225]. The SPD methods, particularly ECAP [226-228] and HPT [229-231], were successfully used to solve the kinetic issues of metal hydrides by mechanochemistry [232]. Moreover, there is a successful attempt to solve the thermodynamic aspects of metal hydrides by using the concepts of binding-energy engineering and ultra-SPD which resulted in the discovery of the first Mg-based alloy for reversible hydrogen storage at room temperature [233]. Similar concepts were employed to discover new high-entropy materials with the capability to store hydrogen at room temperature. The first attempt was synthesizing MgTiVCrFe by SPD which resulted in the BCC + amorphous structure with poor hydrogen storage at room temperature [117]. The second attempt was TiZrCrMnFeNi with the C14 Laves phase structure, which was originally synthesized by SPD but later could be synthesized by conventional arc melting [113]. These studies were later employed to design other high-entropy alloys for low/room-temperature hydrogen storage such as TiZrNbFeNi [234], TiZrNbCrFe [235] and $Ti_xZr_{2-x}CrMnFeNi$ ($x$ = 0.4-1.6) [236].

The high-entropy hydrogen storage materials were designed based on three criteria [113]. (i) The $AB_2$ system is the most appropriate to have low hydrogen binding energy, where A denotes the hydride-forming metals like magnesium, titanium, zirconium, vanadium, and niobium, and B denotes metals with low affinity with the hydrogen atoms like chromium, manganese, iron, cobalt, and nickel. (ii) The average valence electron concentration should be close to 6.4. (iii) The C14 Laves phase has the highest possibility for room-temperature hydrogen storage. The use of these criteria is expected to lead to a hydrogen binding energy slightly lower than 0.1 eV which is a target for reversible room-temperature hydrogen storage, as shown in Fig. 4a [236]. For the $AB_2$-type TiZrCrMnFeNi with a valence electron concentration of 6.4, the CALPHAD calculations confirmed the theoretical stability of the C14 Laves phase structure [113] and first-principles calculations confirmed an appropriate hydrogen binding energy of close to -0.1 eV, as shown in Figs. 4b and 4c [236]. The synthesized TiZrCrMnFeNi alloy with an equiatomic fraction of elements exhibited a C14 Laves phase structure with reversible room-temperature hydrogen storage of 1.7 wt% (30% higher capacity compared to commercial $LaNi_5$), as shown in Fig. 3d [113]. The hydrogenation kinetics of the alloy was also good, it absorbed hydrogen without any need for an activation process, and its hydrogenation rate was very fast, as shown in Fig. 3e [113]. Moreover, the TiZrCrMnFeNi alloy could be handled under an air atmosphere, while hydrogen storage materials should be usually handled under a controlled atmosphere (e.g., in a glove box). Moreover, the examination of cyclic stability, conducted for a non-equiatomic alloy $Ti_{0.4}Zr_{1.6}CrMnFeNi$, confirmed excellent cycling stability for at least 1,000 cycles without any loss in the storage capacity, as shown in Fig. 4f [236]. The room-temperature hydrogen storage capability of $Ti_xZr_{2-x}CrMnFeNi$ ($x$ = 0.4-1.6) alloys resulted in the first experimental application of high-entropy alloys as the anode material of nickel-metal hydride (Ni-MH) batteries, although the discharge capacity of these alloys was not as high as the commercial Ni-MH batteries [237].

In conclusion, the high-entropy alloys, designed by binding-energy engineering and primarily synthesized by SPD, have introduced new candidates for fast and reversible hydrogen storage at room temperature with capacities larger than commercial hydrogen storage materials such as $LaNi_5$. These materials are currently designed by using transition metals to have a C14 Laves crystal structure with a hydrogen-to-metal atomic ratio of 1, but future studies are expected to focus on the design of high-entropy alloys with higher hydrogen-to-metal atomic ratios such as the BCC alloys and on the use of lighter elements. The employment of theoretical concepts is essential for the future design of such high-entropy hydrogen storage materials.

## 6. Photovoltaics

Photovoltaics, which is used in solar cells, is defined as the conversion of light to electric current using semiconductors [238]. Currently, silicon and GaAs show the highest efficiency to generate photocurrent [239,240], but there are significant efforts all over the world to discover new semiconductors for such applications. The high-entropy ceramics, including $TiZrHfNbTaO_{11}$ [126,155] and $TiZrHfNbTaO_6N_3$ [127,154] semiconductors, are one of the newest families of photovoltaic materials synthesized by SPD. These materials were synthesized from elemental metallic powders by HPT and subjected to oxidation and nitriding to produce high-entropy oxides and oxynitrides, respectively. These high-entropy materials were used in the form of thin films by deposition on fluorine-doped tin oxide glass and used photocurrent generation. Fig. 5a shows the appearance of $TiZrHfNbTaO_{11}$ powder [126], Fig. 5b shows its microstructures [126], and Fig. 5c shows its photocurrent generation [155]. It is evident that the oxide can generate electric current under light, although its efficiency is not still comparable to conventional photovoltaic materials. One reason for the low efficiency of this semiconductor is the technical difficulty in fabricating a dense thin film, an issue that needs to be addressed in future studies. Despite this technical difficulty, SPD contributed to the field of photovoltaics by introducing not only high-entropy semiconductors [126,127,154,155] but also some other semiconductors such as high-pressure $TiO_2$-II columbite phase [241] and black $Bi_2O_3$ [242].

Although the efficiency of SPD-synthesized high-entropy ceramic for photocurrent generation is still lower than commercial silicon and GaAs materials, research on this issue is still in its early stages. Future studies should focus not only on the design and synthesis of effective semiconductors for photovoltaics but also on using effective technologies to make dense thin films from these high-entropy semiconductors.

## 7. Photocatalytic water splitting

Photocatalysis is a clean technology to produce hydrogen from water splitting under solar energy in the presence of a light-absorbing catalyst such as $TiO_2$ [243,244]. Despite high scientific interest in photocatalytic hydrogen production, the efficiency of the method is still low for practical applications, and thus, there is a high demand to discover new photocatalysts with high activity [243,244]. A good photocatalyst should have some features such as large light absorbance, appropriate band structure compared to potentials for hydrogen and oxygen production, low bandgap, slow electron-hole recombination rate, a large fraction of active sites on the surface, and high chemical stability [245]. High-entropy photocatalysts were recently introduced as a new family of materials for photocatalytic water spitting [126,127,160]. High-entropy ceramics - which were also used as catalysts for electrocatalytic $O_2$ generation [246-248], chemical oxidation of CO [249-251], operation of lithium-sulfur batteries [252], electrocatalytic $CO_2$ conversion [253], operation of electrochemical capacitors [254], and chemical combustion reactions [255] - have some features that make them attractive as catalysts [256]. (i) They have low free Gibbs energy which gives them high chemical stability. (ii) They have a large fraction of inherent surface defects as active sites for photocatalysis. (iii) The band structure and bandgap of these materials can be adjusted easier than conventional photocatalysts by changing the type or fraction of principal elements.

$TiZrHfNbTaO_{11}$ is the first high-entropy oxide photocatalyst which was synthesized by HPT processing followed by oxidation for photocatalytic hydrogen production [126]. The second material was high-entropy oxynitride photocatalyst $TiZrHfNbTaO_6N_3$ synthesized by a three-step process including HPT processing, oxidation, and high-temperature nitriding [127]. As shown in Fig. 6a, the materials show higher light absorbance compared to all binary oxides and high-entropy oxides in the Ti-Zr-Hf-Nb-Ta system. The material shows high stability for photocatalytic hydrogen production, and it remains active for at least six months (Fig. 6b), while low stability is a general drawback of low-bandgap nitrides and oxynitrides [245]. In the most recent attempt, $TiZrNbTaWO_{12}$ with ten different kinds of heterojunctions (i.e., interphase boundaries which can ease electron-hole separation and migration) was produced by HPT processing followed by oxidation [160]. $TiZrNbTaWO_{12}$ could produce oxygen from water under visible light (Fig. 6c), while the activity of most photocatalysts is limited to the ultraviolet (UV) range of sunlight [243,244].

In conclusion, high-entropy photocatalysts, which were first synthesized by SPD, show high potential for water splitting and it is expected that a wider range of compositions and different synthesis methods will be employed in the future to utilize these materials. The key point in this respect is the design of active photocatalysts through the use of theoretical concepts. Moreover, some technologies should be employed to increase the surface area of these high-entropy photocatalysts by producing nanopowders.

## 8. $CO_2$ photoreduction

$CO_2$ emission from human activity has made the serious problem of global warming, an issue that is expected to influence the lives of many people in the 21st century [257]. In addition to the utilization of zero-$CO_2$ emission fuels such as hydrogen, the capture of $CO_2$ and its conversion to reactive gases such as CO or value-added hydrocarbons and fuels is another solution to deal with the $CO_2$ emission issues [258]. Photoreduction of $CO_2$ under sunlight and in the presence of a semiconductor catalyst is the cleanest technology for such conversions, but the field is rather new and needs significant improvement particularly by discovering new catalysts [259]. The SPD method was effectively used to improve $CO_2$ photoreduction by inducing high-pressure phases and oxygen vacancies in typical photocatalysts such as $TiO_2$ [260] and $BiVO_4$ [261]. The introduction of high-entropy oxides and oxynitrides, synthesized by SPD, is another successful attempt in this regard [154,155].

The band structure and the position of the valence band top and the conduction band bottom for $TiZrHfNbTaO_{11}$ and $TiZrHfNbTaO_6N_3$, which were originally designed and synthesized for photocatalytic water splitting [126,127], satisfy the energy requirements for $CO_2$ photoreduction, as shown in Fig. 7a [154]. While the high-entropy oxide has a bandgap of 3 eV which is similar to $TiO_2$, the high-entropy oxynitride shows a low bandgap of 1.6 eV. In addition to low bandgap and appropriate band structure, physical adsorption and chemisorption of $CO_2$ on high-entropy catalysts, particularly on $TiZrHfNbTaO_6N_3$, is higher than $TiO_2$, as shown in Fig. 7b using diffuse reflectance infrared Fourier transform spectroscopy [154]. Moreover, photoluminescence measurements suggest lower electron-hole recombination in this oxynitride compared to relevant high-entropy oxide and $TiO_2$ [154]. $TiZrHfNbTaO_{11}$ shows high activity for $CO_2$ photoreduction compared to P25 $TiO_2$ which is a benchmark catalyst, while $TiZrHfNbTaO_6N_3$ exhibits even higher $CO_2$ photoreduction activity, as shown in Fig. 7c [154].

These preliminary studies confirm the contribution of SPD in introducing some of the most active catalysts for $CO_2$ photoreduction by using the concept of high-entropy ceramics. Despite significant success in the experimental development of such ceramics, theoretical calculations are needed to clarify the mechanisms underlying their high activity. CALPHAD, density functional theory calculations, molecular dynamics simulations, and multiscale modeling are some approaches that can shed light on understanding the mechanism of activity of these new catalysts and their mechanochemistry by SPD processing [262-264].

## 9. Concluding remarks and future outlook

Superfunctional high-entropy alloys and ceramics, with properties superior to the normal functions of engineering materials, can be developed by processing or synthesizing through severe plastic deformation (SPD). The SPD has led to various superfunctional properties including (i) ultrahigh hardness, (ii) high strength with good hydrogen embrittlement resistance, (iii) high strength, and biocompatibility with low elastic modulus, (iv) fast and reversible room-temperature hydrogen storage, (v) photocurrent generation, (vi) photocatalytic water splitting, and (vii) $CO_2$ photoreduction. These findings not only suggest new applications for SPD but also introduce severely deformed high-entropy alloys and ceramics as a new family of superfunctional materials with high potential for various applications. The application of severely deformed high-entropy materials is expected to be expanded by considering the current demands to discover advanced energy materials with higher efficiency.

## Acknowledgments

The author P.E. thanks the Hosokawa Powder Technology Foundation of Japan for a grant. The author K.E. was supported by the MEXT, Japan through Grants-in-Aid for Scientific Research on Innovative Areas (JP19H05176 & JP21H00150), and in part by the MEXT, Japan through Grant-in-Aid for Challenging Research Exploratory (JP22K18737).

## Appendix

Table A1 summarizes the publications on the application of SPD as a processing or synthesis tool to medium-entropy materials (containing four principal elements) and high-entropy materials (containing at least five principal elements) [66-169].

Table A1. Summary of publications about severe plastic deformation of medium- and high-entropy alloys and ceramics in chronological order.

| Composition | Crystal Structure | SPD Method | SPD Role | Properties or Features | References |
|---|---|---|---|---|---|
| AlCoCrCuFeNi | BCC + FCC | MDF | Processing | Superplasticity | Kuznetsov *et al.* (2013) [66] |
| AlFeMgTiZn | Multiphases | ECAP | Processing | Mechanical properties | Hammond *et al.* (2014) [67] |
| CoCrFeHMnNi | FCC | HPT | Processing | Mechanical properties and thermal stability | Schuh *et al.* (2015) [68] |
| $Al_{0.3}CoCrFeNi$ | FCC | HPT | Processing | Hardening by straining and annealing | Tang *et al.* (2015) [69] |
| CoCrFeNiMn | FCC | HPT | Processing | Nanomechanical behavior | Lee *et al.* (2015) [70] |
| $Al_{0.3}CoCrFeNi$ | FCC | HPT | Processing | Microstructural features | Tang *et al.* (2015) [71] |
| $Al_{0.1}CoCrFeNi$ | FCC | HPT | Processing | Microstructural features | Yu *et al.* (2016) [72] |
| $Al_{0.3}CoCrFeNi$ | FCC | HPT | Processing | Annealing-induced phase transformation | Tang et al. (2016) [73] |
| CoCrFeMnNi | FCC | HPT | Processing | Nanoindentation creep behavior | Lee *et al.* (2016) [74] |
| CoCrFeNiMn | FCC | HPT | Processing | Annealing effect on mechanical properties | Shahmir *et al.* (2016) [75] |
| $Al_{0.3}Cu_{0.5}CoCrFeNi$ | FCC | HPT | Processing | Strain-induced homogenization | Yuan *et al.* (2016) [76] |
| $Ti_{35}Zr_{27.5}Hf_{27.5}Nb_5Ta_5$ | HCP | HPT | Processing | Microstructure and hardness | Heczel *et al.* (2017) [77] |
| CrMnFeCoNi | FCC | HPT | Processing | Phase stability examination by nanoindentation | Maier-Kiener *et al.* (2017) [78] |
| CoCrFeNiMn | FCC | HPT | Processing | Superplasticity | Shahmir *et al.* (2017) [79] |
| AlTiVNb | BCC | HPT | Processing | Phase transformation | Schuh *et al.* (2017) [80] |
| CrMnFeCoNi | FCC | HPT | Processing | Microstructure and texture evolution | Skrotzki *et al.* (2017) [81] |
| CoCrFeNiMn | FCC | ECAP | Processing | Microstructural features | Shahmir *et al.* (2017) [82] |
| $(FeNiCoCu)_{1-x}Ti_xAl_x$ | FCC | HPT | Processing | Tensile properties | Zheng *et al.* (2017) [83] |
| $Ti_{35}Zr_{27.5}Hf_{27.5}Nb_5Ta_5$ | Orthorhombic | HPT | Processing | Lattice defects | Heczel *et al.* (2017) [84] |
| CoCrFeMnNi | FCC | HPT | Processing | Lattice defects and hardness | Heczel *et al.* (2017) [85] |
| CoCrFeMnNi | FCC + σ | HPT | Processing | Phase stability | Park *et al.* (2017) [86] |
| FeCoCrNi | FCC | HPT | Processing | Amorphization and twining | Wu *et al.* (2017) [87] |
| $Co_{20}Cr_{26}Fe_{20}Mn_{20}Ni_{14}$ | HCP | HPT | Processing | Phase transformation | Moon *et al.* (2017) [88] |
| CoCrFeMnNi | FCC | HPT | Processing | Annealing effect on plastic flow | Lee *et al.* (2017) [89] |
| $Al_{0.5}CoCrFeMnNi$ | FCC + B2 | HPT | Processing | Microstructure and phase transformation | Reddy *et al.* (2017) [90] |
| TiZrNbHfTa | BCC | HPT | Processing | Thermodynamic instability | Schuh *et al.* (2018) [91] |
| $Co_{20}Cr_{20}Fe_{20}Mn_{20}Ni_{20}$ | FCC | HPT | Processing | Micro-scale mechanical behavior | Kawasaki *et al.* (2018) [92] |
| CoCrFeNiMn | FCC | HPT | Processing | Superplasticity | Shahmir *et al.* (2018) [93] |
| CoCrFeMnNi | FCC | Caliber Rolling | Processing | Microstructural features | Won *et al.* (2018) [94] |
| CoCrFeMnNi | FCC | HPT | Processing | Phase stability and strengthening | Shahmir *et al.* (2018) [95] |
| CoCrFeNiMn | FCC | HPT | Processing | Superplasticity | Shahmir *et al.* (2018) [96] |
| Carbon-doped $CrFe_2NiMnV_{0.25}$ | FCC + Carbide | HPT | Processing | Carbon and annealing effect on hardness | Shahmir *et al.* (2018) [97] |
| HfNbTaTiZr | BCC | HPT | Processing | Lattice defects | Lukáča *et al.* (2018) [98] |
| HfNbTaTiZr | BCC | HPT | Processing | Mechanical properties | Čížek *et al.* (2018) [99] |
| $CoCrFeNiNb_x$ | FCC | HPT | Processing | Dislocation structure | Maity *et al.* (2018) [100] |
| CoCrFeMnNi | FCC | HPT | Cryogenic Processing | Microstructure and mechanical properties | Zherebtsov *et al.* (2018) [101] |
| $AlNbTiVZr_{0.5}$ | B2 | HPT | Processing | Microstructural and hardness | Stepanov *et al.* (2018) [102] |
| $Al_{0.3}CrFeCoNi$ | BCC + FCC | HPT | Processing | Deformation behavior | Qiang *et al.* (2018) [103] |
| CoCrFeMnNi | FCC | HPT | Synthesis | Activation energy for plastic flow | Lee *et al.* (2018) [104] |
| CoCrFeMnNi | FCC | HPT | Synthesis | Microstructure and hardness | Kilmametov *et al.* (2019) [105] |
| $Hf_{25}Nb_{25}Ti_{25}Zr_{25}$ | BCC | HPT | Processing | Microstructure and hardness | Gubicza *et al.* (2019) [106] |
| CoCrFeNi | FCC | HPT | Processing | Microstructure and hardness | Gubicza *et al.* (2019) [107] |
| CoCuFeMnNi | FCC | HPT | Monotonic and Cyclic Processing | Microstructure and hardness | Sonkusare *et al.* (2019) [108] |
| CrMnFeCoNi | FCC | HPT | Processing | Ductility | Schuh *et al.* (2019) [109] |
| CoCrFeNiMn | FCC | HPT | Processing | Nanoindentation creep behavior | Zhou *et al.* (2019) [110] |
| $V_{10}Cr_{15}Mn_5Fe_{35}Co_{10}Ni_{25}$ | FCC | HPT | Processing | Superplasticity | Nguyen *et al.* (2019) [111] |
| HfNbTaTiZr | BCC | HPT | Processing | Microstructure and mechanical properties | Málek *et al.* (2019) [112] |
| TiZrCrMnFeNi | C14 + B2 | HPT | Processing | Hydrogen storage | Edalati *et al.* (2020) [113] |
| $V_{10}Cr_{15}Mn_5Fe_{35}Co_{10}Ni_{25}$ | FCC | HPT | Processing | Grain size effect on deformation mechanism | Asghari-Rad *et al.* (2020) [114] |
| CoCrFeNiMn | FCC | HPT | Cryogenic Processing | Microstructure and strength | Podolskiy *et al.* (2020) [115] |
| $V_{10}Cr_{15}Mn_5Fe_{35}Co_{10}Ni_{25}$ | FCC | HPT | Processing | Mechanical properties | Asghari-Rad *et al.* (2020) [116] |
| MgVTiCrFe | BCC | HPT | Processing | Hydrogen storage | de Marco *et al.* (2020) [117] |
| CoCuFeMnNi | FCC | HPT | Processing | Microstructure, texture and mechanical properties | Sonkusare *et al.* (2020) [118] |
| CoCrFeNi | FCC | HPT | Processing | Strain-rate sensitivity | Zhao *et al.* (2020) [119] |
| $Fe_{20}Mn_{20}Ni_{20}Co_{20}Cr_{20}$ | FCC | HPT | Processing | Hydrogen diffusion | Belo *et al.* (2020) [120] |
| $Fe_{41}Mn_{25}Ni_{24}Co_8Cr_2$ | FCC | Differential Rolling | Processing | Tensile properties | Jeong *et al.* (2020) [121] |
| CrMnFeCoNi | FCC + HCP | HPT | Processing | Microstructure, texture and strength | Skrotzki *et al.* (2020) [122] |
| TiAlFeCoNi | BCC + $L2_1$ | HPT | Processing | Biocompatibility | Edalati *et al.* (2020) [123] |
| CrFeCoNi | FCC | ECAP | Processing | Microstructure | Rymer *et al.* (2020) [124] |
| HfNbTiZr | BCC + HCP | HPT | Processing | Thermal stability | Hung *et al.* (2020) [125] |
| $TiZrNbHfTaO_{11}$ | Monoclinic + Orthorhombic | HPT | Synthesis | Photocatalytic $H_2$ production | Edalati *et al.* (2020) [126] |
| $TiZrNbHfTaO_6N_3$ | FCC + Monoclinic | HPT | Synthesis | Photocatalytic $H_2$ production | Edalati *et al.* (2021) [127] |
| CoCrNiFeMn | FCC | ECAP | Processing | Low-cycle fatigue | Picak *et al.* (2021) [128] |
| CoCrFeNiMn | FCC + BCC | HPT | Processing | Phase transformation | Shahmir *et al.* (2021) [129] |
| Carbon-doped CoCrFeMnNi | FCC | HPT | Processing | Carbon effect on microstructure | Lu *et al.* (2021) [130] |
| $Co_{17.5}Cr_{12.5}Fe_{55}Ni_{10}Mo_5$ | FCC | HPT | Processing | Strengthening by precipitates and nanograins | Kwon *et al.* (2021) [131] |
| $Fe_{40}Mn_{40}Co_{10}Cr_{10}$ | HCP + FCC | HPT | Processing | Plasticity and non-basal slip | Chandan *et al.* (2021) [132] |
| CoCrFeMnNi | FCC | ECAP | Processing | Twinning and phase transformation | Picak *et al.* (2021) [133] |
| TiNbZrTaHf | BCC | HPT | Synthesis | Biocompatibility | González-Masís *et al.* (2021) [134] |
| $CoCrFeNiNb_x$ | FCC | HPT | Processing | Strength and elasticity | Maity *et al.* (2021) [135] |
| Carbon-doped AlTiFeCoNi | BCC + FCC | HPT | Processing | Ultrahigh hardness and phase transformation | Edalati *et al.* (2021) [136] |
| AlCrFeCoNiNb | BCC + C14 | HPT | Processing | Ultrahigh hardness | Edalati *et al.* (2021) [137] |
| CoCrFeMnNi | FCC | HPT | consolidation | Mechanical properties | Asghari-Rad *et al.* (2021) [138] |
| $Fe_{40}Mn_{40}Co_{10}Cr_{10}$ | FCC | HPT | Processing | Strain-induced phase transformation and plasticity | Sathiyamoorthi *et al.* (2021) [139] |
| $(CrMnFeCo)_xNi_{1-x}$ | FCC | HPT | Processing | Microstructural saturation | Keil *et al.* (2021) [140] |
| CoCrFeNi | FCC | HPT | Processing | Micromechanical properties | Zhao *et al.* (2021) [141] |
| CoCrFeMnNi-based Composite | FCC | HPT | synthesis | Microstructure and mechanical properties | Karthik *et al.* (2021) [142] |
| CoCrFeMnNi + HfNbTaTiZr Composite | FCC + BCC | HPT | Synthesis | Microstructure | Taheriniya *et al.* (2021) [143] |
| $AlCrFe_2Ni_2$ | FCC | HPT | Processing | Phase transformation | Liu *et al.* (2021) [144] |
| Various Materials | Review paper on synthesis of high-entropy alloys and ceramics by HPT | | | | Edalati *et al.* (2021) [65] |
| CoCrFeNiMn | FCC | HPT | Processing | Micromechanical properties | Rusakova *et al.* (2021) [145] |
| CoCrFeNi | FCC | HPT | Processing | Annealing effect on microstructure | Hung *et al.* (2021) [146] |
| CoCrFeMnNi | FCC | HPT | consolidation | Microstructural features | Asghari-Rad *et al.* (2021) [147] |
| Various Alloys | Review paper regarding the impact of HPT on mechanical properties of high-entropy alloys | | | | Li *et al.* (2022) [148] |
| $Al_{0.5}CoCrFeMnNi$ | FCC | HPT | Processing | Superplasticity | Nguyen *et al.* (2022) [149] |
| CoCrFeMnNi | FCC | HPT | Processing | Corrosion resistance | Shimizu *et al.* (2022) [150] |
| FeCoNiCu | FCC | HPT | Processing | Mechanical properties and thermal stability | Zhang *et al.* (2022) [151] |
| $Fe_{40}Mn_{40}Co_{10}Cr_{10}$ | FCC | HPT | Processing | Nickel effect on microstructure | Chandan *et al.* (2022) [152] |
| $Al_{0.7}CoCrFeNi$ | FCC + BCC | HPT | Synthesis | Hydrogen effect on strengthening | Gao *et al.* (2022) [153] |
| $TiZrNbHfTaO_6N_3$ | FCC + Monoclinic | HPT | Synthesis | $CO_2$ photoreduction | Akrami *et al.* (2022) [154] |
| $TiZrNbHfTaO_{11}$ | Monoclinic + Orthorhombic | HPT | Synthesis | $CO_2$ photoreduction | Akrami *et al.* (2022) [155] |
| CrMnFeCoNi | FCC | HPT | Processing | Mechanical properties and thermal stability | Keil *et al.* (2022) [156] |
| CoCrFeNi | FCC | HPT | Synthesis | Tensile properties | Son *et al.* (2022) [157] |
| $FeNi_2CoMo_{0.2}V_{0.5}$ | FCC | HPT | Processing | Thermal stability | Liang *et al.* (2022) [158] |
| $Al_{0.3}CoCrNi$ | FCC + B2 + σ | HPT | Processing | Superplasticity | Nguyen *et al.* (2022) [159] |
| $TiZrNbTaWO_{12}$ | Multiphases | HPT | Synthesis | Photocatalytic $O_2$ production | Edalati *et al.* (2022) [160] |
| CoCrFeNi | FCC | HPT | Processing | Thermal stability | Hung *et al.* (2022) [161] |
| Various Alloys | Review on phase transformations of severely deformed high-entropy alloys | | | | Straumal *et al.* (2022) [162] |
| AlFeCoNiCu | FCC + BCC | HPT | Processing | Lattice defects and microhardness | Edalati *et al.* (2022) [163] |
| MoNbTaTiVZr | BCC + BCC | HPT | Processing | Strain hardening | Duan *et al.* (2022) [164] |
| CrMnFeCoNi | | HPT | | Hydrogen embrittlement | Mohammadi *et al.* (2022) [165] |
| FeCoNiMn | FCC | HPT | Processing | Homogenization effect on hardness | Alijani *et al.* (2022) [166] |
| CrMnFeCoNi | | HPT | | Hydrogen embrittlement | Mohammadi *et al.* (2022) [167] |
| $Co_{25-x}Cr_{25}Fe_{25}Ni_{25}C_x$ | FCC | HPT | Processing | Micromechanical properties | Levenets *et al.* (2022) [168] |
| $Al_{0.1}CoCrFeNi$ | FCC | ECAP | Processing | Microstructure and mechanical properties | Bian et al. (2022) [169] |

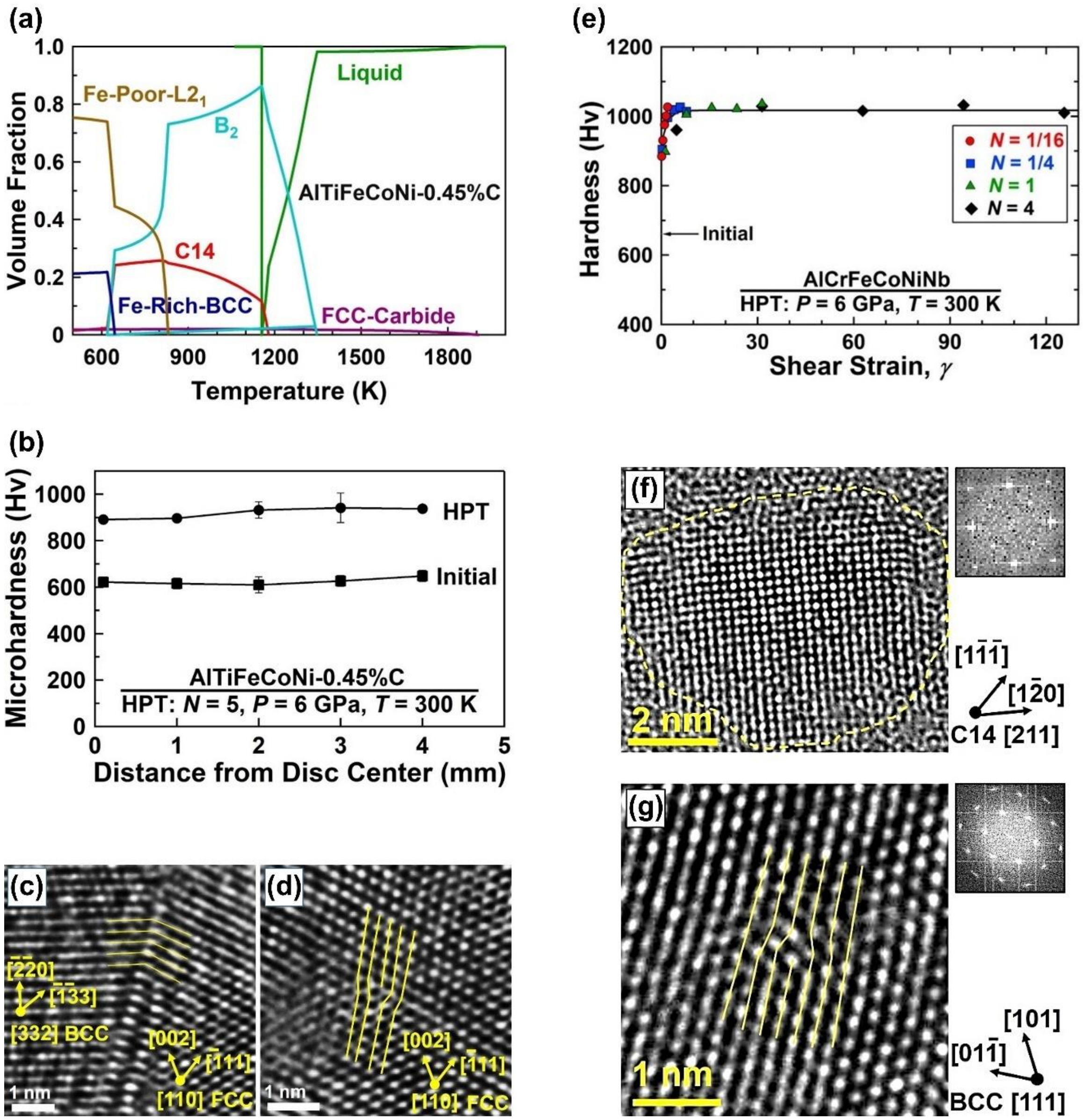

Fig. 1 Development of ultrahard high-entropy alloys by severe plastic deformation. **a** Phase diagram of AlTiFeCoNi doped with 0.45 wt% of carbon calculated by CALPHAD. Reproduced with permission from Ref. [136]. Copyright 2021, Elsevier. **b** Hardness of carbon-doped AlTiFeCoNi before and after HPT processing. Reproduced with permission from Ref. [136]. Copyright 2021, Elsevier. **c, d** High-resolution lattice images of HPT-processed carbon-doped AlTiFeCoNi showing the presence of interphases and dislocations. Reproduced with permission from Ref. [136]. Copyright 2021, Elsevier. **e** Hardness against shear strain for AlCrFeCoNiNb processed by HPT for various turns, *N*. Reproduced with permission from Ref. [137]. Copyright 2021, Elsevier. **f, g** Lattice images and corresponding diffractograms for AlCrFeCoNiNb processed by HPT for *N* = 4 showing the presence of nanograins and dislocations. Reproduced with permission from Ref. [137]. Copyright 2021, Elsevier.

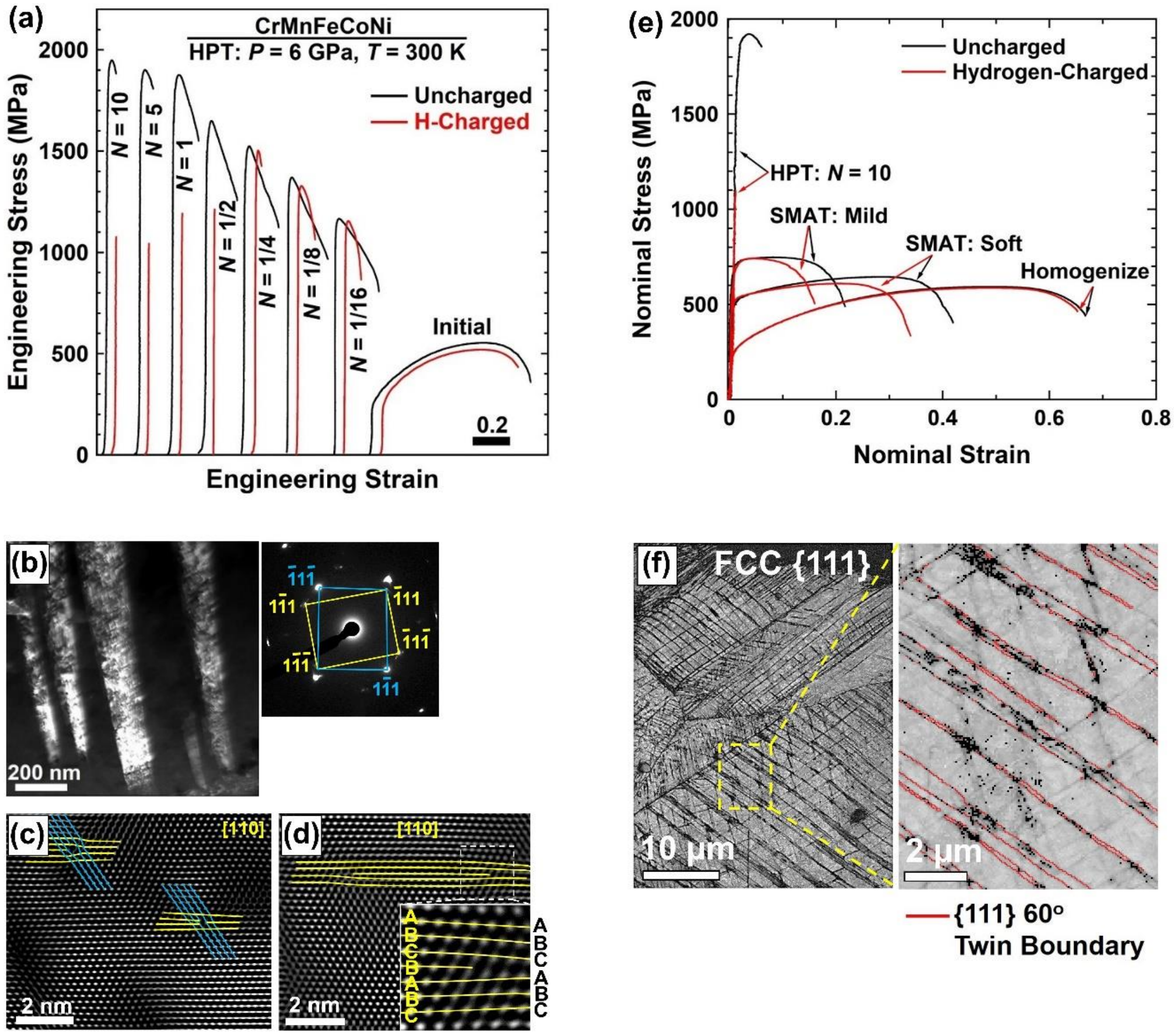

Fig. 2 Development of high yield strength and good resistance to hydrogen embrittlement in high-entropy alloys by severe plastic deformation. **a** Tensile stress-strain curves before and after hydrogen charging for Cantor alloy CrMnFeCoNi processed by HPT for various rotations, *N*. Reproduced with permission from Ref. [167]. Copyright 2022, Elsevier. **b** dark-field image and corresponding selected area electron diffraction patterns for CrMnFeCoNi processed with HPT for *N* = 1/16, showing the presence of twins. Reproduced with permission from Ref. [167]. Copyright 2022, Elsevier. **c, d** High-resolution lattice images of CrMnFeCoNi processed by HPT for *N* = 1/4, showing the presence of Lomer-Cottrell locks and D-Frank partial dislocations. Reproduced with permission from Ref. [167]. Copyright 2022, Elsevier. **e** Tensile stress-strain curves before and after hydrogen charging for CrMnFeCoNi processed by homogenization, soft SMAT (40 μm ultrasonic amplitude for 30 s), mild SMAT (60 μm ultrasonic amplitude for 60 s) and intense HPT (*N* = 10). Reproduced with permission from Ref. [165]. Copyright 2022, Elsevier. **f** Band contrast image and {111} twin boundary mapping by electron backscatter diffraction in scanning electron microscopy for CrMnFeCoNi processed by SMAT under mild conditions. Reproduced with permission from Ref. [165]. Copyright 2022, Elsevier.

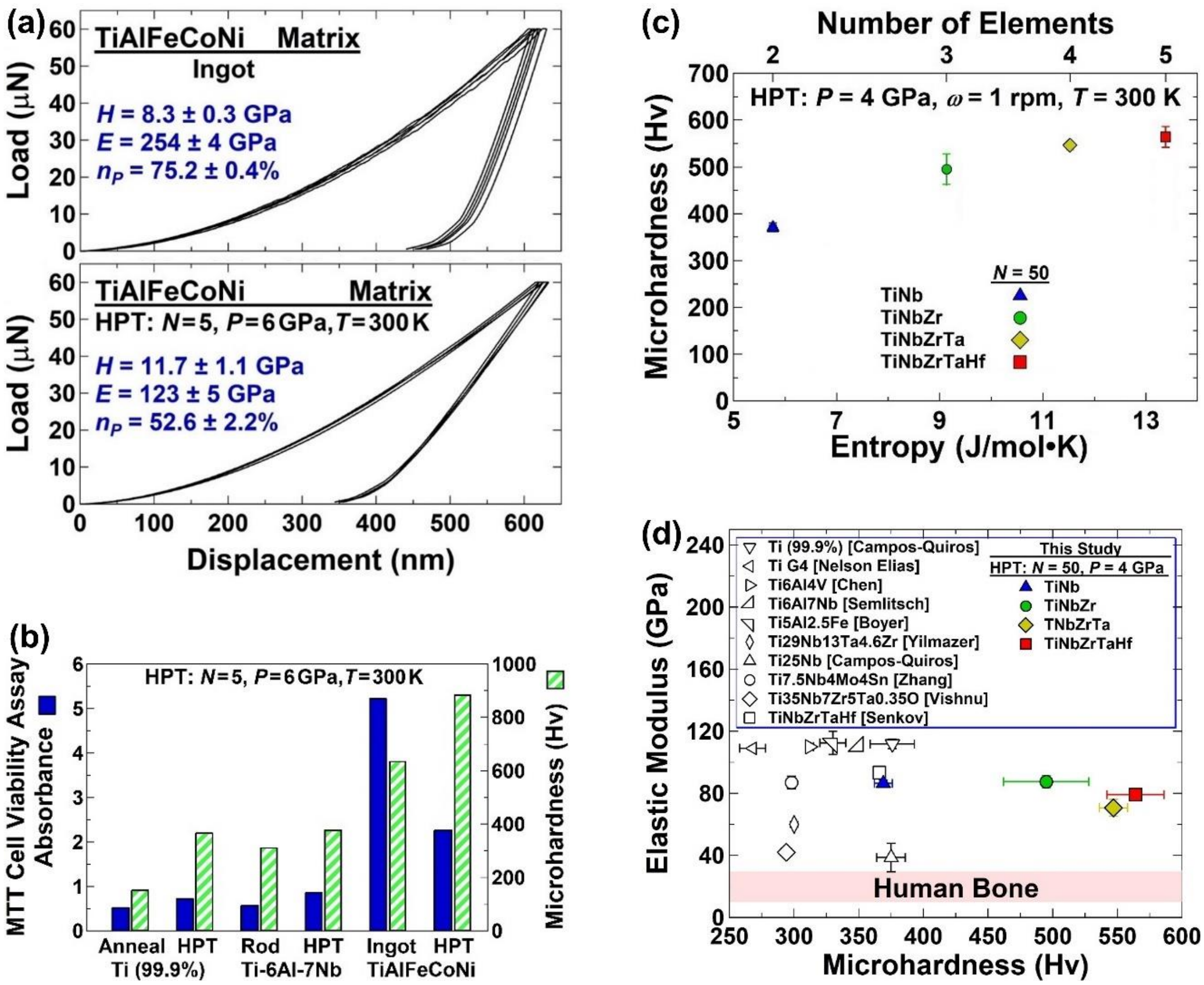

Fig. 3 Development of high-entropy alloys with high strength, high biocompatibility and low elastic modulus by severe plastic deformation. **a** Nanoindentation load against displacement for TiAlFeCoNi after arc melting (ingot) and HPT processing. Reproduced with permission from Ref. [123]. Copyright 2020, Elsevier. **b** Biocompatibility, examined by MTT cell viability assay, and hardness of ingot and HPT-processed TiAlFeCoNi samples in comparison with pure titanium after annealing and HPT processing and Ti-6Al-7Nb (wt.%) after extrusion (rod) and HPT processing. Reproduced with permission from Ref. [123]. Copyright 2020, Elsevier. **c** Hardness against configurational entropy and the number of elements for TiNb, TiNbZr, TiNbZrTa and TiNbZrTaHf synthesized by HPT. Reproduced with permission from Ref. [134]. Copyright 2021, Elsevier. **d** Elastic modulus and hardness of TiNb, TiNbZr, TiNbZrTa and TiNbZrTaHf synthesized by HPT in comparison with some conventional biomaterials and human bone. Reproduced with permission from Ref. [134]. Copyright 2021, Elsevier.

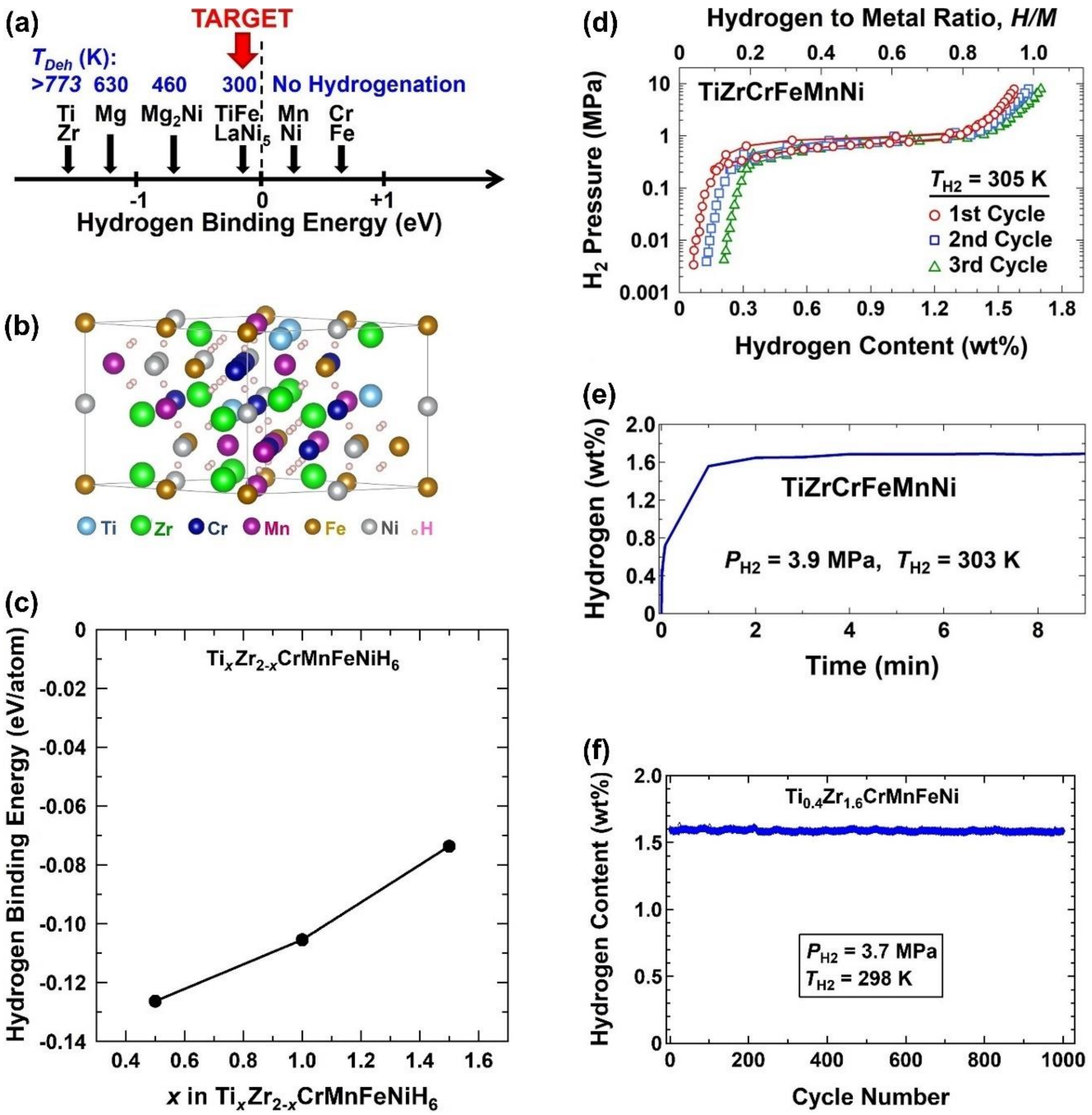

Fig. 4 Design of high-entropy alloys with fast and reversible hydrogen storage at room temperature by binding-energy engineering. **a** Illustration of the concept of hydrogen-binding energy to design room-temperature hydrogen storage materials. Reproduced with permission from Ref. [233]. Copyright 2020, Elsevier. **b** Superlattice of high-entropy hydride $Ti_{0.4}Zr_{1.6}CrMnFeNiH_6$. Reproduced with permission from Ref. [236]. Copyright 2022, Elsevier. **c** Hydrogen binding energy versus titanium content in $Ti_xZr_{2-x}CrMnFeNiH_6$ calculated by first-principles calculations. Reproduced with permission from Ref. [236]. Copyright 2022, Elsevier. **d** Hydrogen pressure-temperature-composition isotherms at room temperature for TiZrCrMnFeNi. Reproduced with permission from Ref. [113]. Copyright 2020, Elsevier. **e** Hydrogen content versus hydrogenation time for TiZrCrMnFeNi at room temperature. Reproduced with permission from Ref. [113]. Copyright 2020, Elsevier. **f** Hydrogen storage content during cycling test for $Ti_{0.4}Zr_{1.6}CrMnFeNi$ at room temperature. Reproduced with permission from Ref. [236]. Copyright 2022, Elsevier.

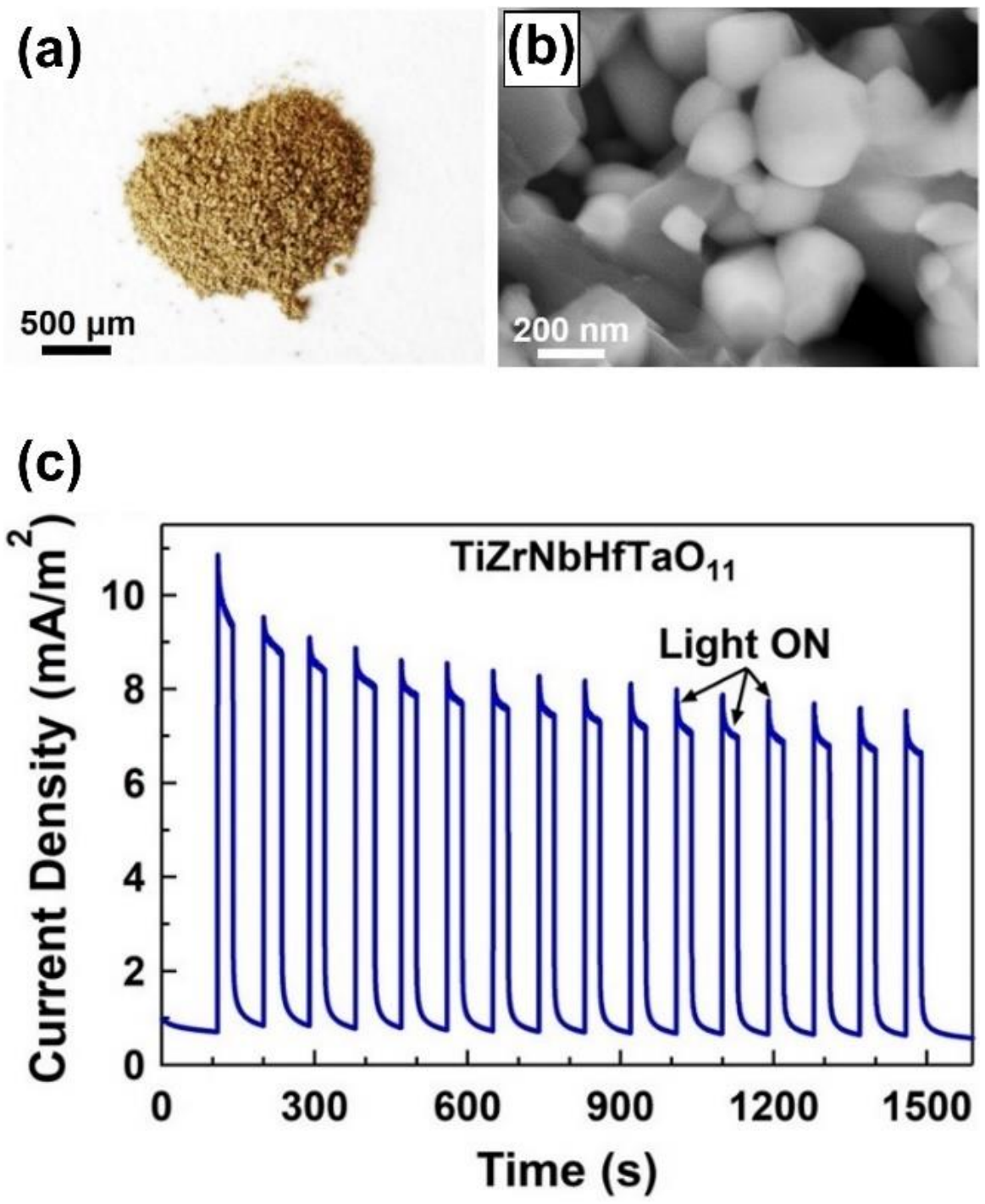

Fig. 5 High-entropy ceramic semiconductors as photovoltaic materials. **a, b** Appearance **b** microstructure visualized by scanning electron microscopy of $TiZrZrNbTaO_{11}$ synthesized by HPT followed by oxidation. Reproduced with permission from Ref. [126]. Copyright 2020, RCS. **c** Photocurrent generation on $TiZrZrNbTaO_{11}$. Reproduced with permission from Ref. [155]. Copyright 2022, Elsevier.

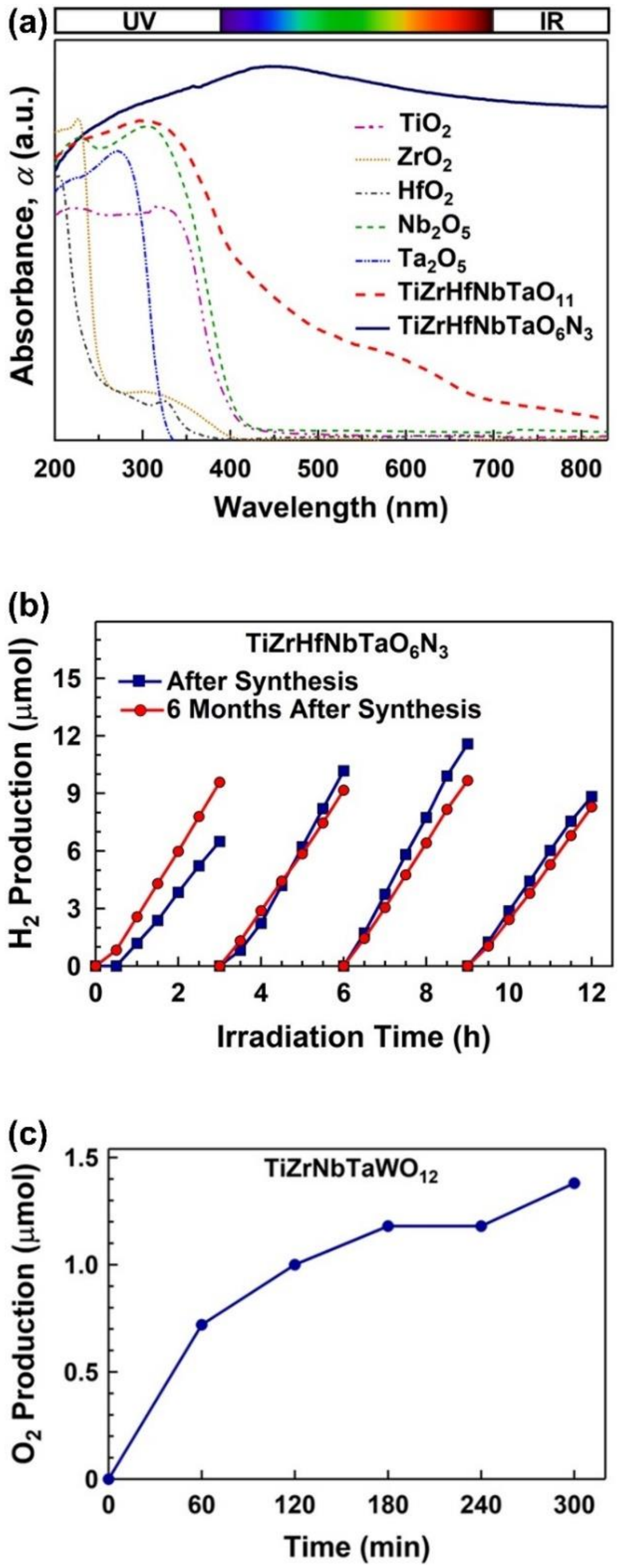

Fig. 6 Photocatalytic water splitting on high-entropy oxides and oxynitrides synthesized by severe plastic deformation. **a** UV-vis light absorbance spectra for $TiZrHfNbTaO_{11}$ and $TiZrHfNbTaO_6N_3$ in comparison with binary oxides. Reproduced with permission from Ref. [127]. Copyright 2021, RCS. **b** Photocatalytic hydrogen production under UV light on $TiZrHfNbTaO_6N_3$ for four cycles after synthesis and after six-month storage. Reproduced with permission from Ref. [127]. Copyright 2021, RCS. **c** Photocatalytic oxygen production on $TiZrNbTaWO_{12}$ with multiple heterojunctions under visible light. Reproduced with permission from Ref. [160]. Copyright 2022, Elsevier.

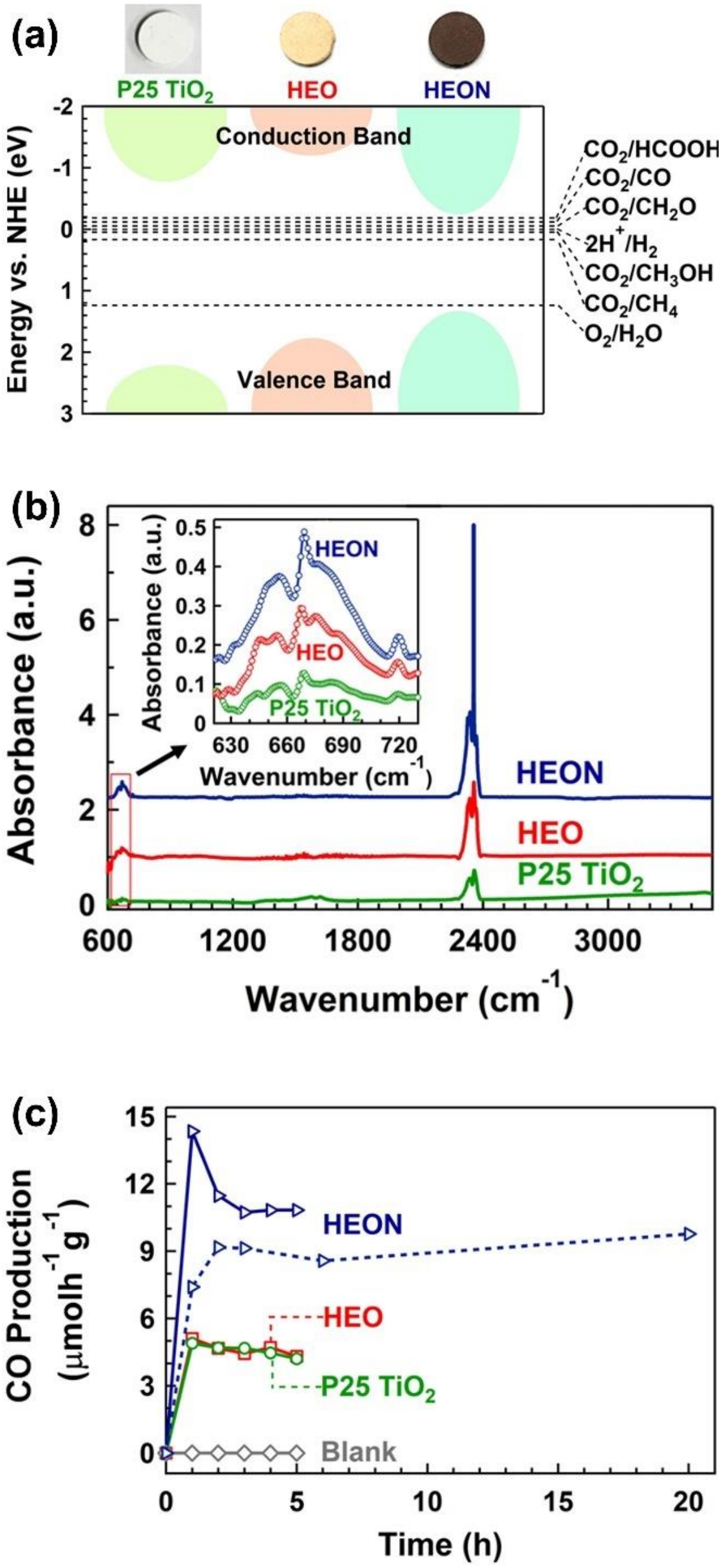

Fig. 7 $CO_2$ photoreduction to CO on high-entropy oxides and oxynitrides synthesized by severe plastic deformation. **a** Electronic band structure and the appearance of $TiZrHfNbTaO_{11}$ and $TiZrHfNbTaO_6N_3$ in comparison with P25 $TiO_2$. Reproduced with permission from Ref. [154]. Copyright 2022, Elsevier. **b** Diffuse reflectance infrared Fourier transform spectra for $TiZrHfNbTaO_{11}$ and $TiZrHfNbTaO_3N_6$ in comparison with P25 $TiO_2$ in which peaks at 665 $cm^{-1}$ and 2350 $cm^{-1}$ correspond to $CO_2$ chemisorption and physisorption on the surface of catalyst. Reproduced with permission from Ref. [154]. Copyright 2022, Elsevier. **c** $CO_2$ photoreduction to CO during UV light illumination time on $TiZrHfNbTaO_{11}$ and $TiZrHfNbTaO_3N_6$ in comparison with P25 $TiO_2$. Reproduced with permission from Ref. [154]. Copyright 2022, Elsevier.